\documentclass[nofootinbib,aps,a4paper,letterpaper,superscriptaddress,
twocolumn,times,eqsecnum]{revtex4}
\usepackage{amsmath}
\usepackage{amsfonts}
\usepackage{booktabs}
\usepackage{multirow}
\usepackage{siunitx}
\usepackage{adjustbox}
 \pdfoutput=1
\usepackage{graphicx}
\usepackage{color}
\usepackage{braket}
\usepackage{dcolumn}
\usepackage{bm,url}
\usepackage[linktocpage]{hyperref}
\usepackage{subfigure}
\usepackage{amsfonts}
\usepackage[usenames,dvipsnames,svgnames]{xcolor}  
\usepackage{hyperref}
\definecolor{oxfordblue}{rgb}{0.0, 0.13, 0.28}
\definecolor{burgundy}{rgb}{0.5, 0.0, 0.13}
\definecolor{darkolivegreen}{rgb}{0.33, 0.42, 0.18}
\definecolor{darkblue}{rgb}{0,0,0.5}
\definecolor{richcarmine}{rgb}{0.84, 0.0, 0.25}
\definecolor{darkblue}{rgb}{0,0,0.5}
\definecolor{bluer}{rgb}{0.00,0.50,0.75}{}
\hypersetup{colorlinks=true, citecolor=red, linkcolor=blue,
 urlcolor = magenta, filecolor=magenta}

\let\oldsqrt\sqrt
\def\sqrt{\mathpalette\DHLhksqrt}
\def\DHLhksqrt#1#2{%
\setbox0=\hbox{$#1\oldsqrt{#2\,}$}\dimen0=\ht0
\advance\dimen0-0.2\ht0
\setbox2=\hbox{\vrule height\ht0 depth -\dimen0}%
{\box0\lower0.4pt\box2}}

\newcommand{\beq}{\begin{equation}}
\newcommand{\eeq}{\end{equation}}
\newcommand{\bea}{\begin{equation}\begin{aligned}}
\newcommand{\eea}{\end{aligned}\end{equation}}

\newlength{\wsingfig}
\newlength{\wdblefig}
\newlength{\wquadfig}
\newlength{\wtriplefig}
\begin{document}

\title{On primordial black hole mass growth through radiative absorption}

\author{Dimitris S. Kallifatides}
\email{dkallifatides@hotmail.com}
\affiliation{Ministry of Education, Marousi, 15180, Athens, Greece}
\affiliation{National Observatory of Athens, Lofos Nymfon, 11852 Athens,
Greece}

\author{Theodoros Papanikolaou}
\email{papaniko@upatras.gr}
\affiliation{Laboratory of Theoretical and Computational Physics, Department of
Physics, University of Patras, 26504 Patras, Greece}

\author{Emmanuel N. Saridakis}
\email{msaridak@noa.gr}
\affiliation{National Observatory of Athens, Lofos Nymfon, 11852 Athens,
Greece}
\affiliation{Departamento de Matem\'{a}ticas, Universidad Cat\'{o}lica del
Norte,
Avda.
Angamos 0610, Casilla 1280 Antofagasta, Chile}
\affiliation{CAS Key Laboratory for Researches in Galaxies and Cosmology,
Department of Astronomy, University of Science and Technology of China, Hefei,
Anhui 230026, P.R. China}

\begin{abstract}

Whether primordial black holes (PBHs) can grow substantially through radiative absorption in
the early Universe remains an important question, particularly in connection with the origin of su-
permassive black holes. If the primordial plasma is treated as an unperturbed
thermal bath, PBHs formed with a horizon-mass fraction $f_H\gtrsim0.2$ are
predicted to grow without bound. We revisit this problem, distinguishing the
tightly coupled plasma, which is accreted hydrodynamically, from free-streaming
neutrinos, which are captured kinetically, accounting also for the Misner-Sharp energy
balance to formulate local radiation depletion. Interestingly, the ingoing
neutrino flux is found to be not locally depleted, whereas the energy deficit created by
fluid accretion can only be redistributed within the sound horizon. With this
causal depletion, we show that no self-similar growth is possible if the accessible reservoir
is smaller than $1.4$ sound horizons, while the runaway pole at
$f_H\gtrsim0.2$ is replaced by a finite amplification. In particular, in the regime of validity of 
quasi-stationary capture, namely for $f_H\lesssim0.25$, radiative
absorption increases the PBH mass by at most a factor of $8$. These results
reveal the crucial role of local energy availability in regulating PBH evolution and
show that substantial PBH mass growth cannot be inferred from radiative
absorption alone.
\end{abstract}

\maketitle

\section{Introduction}
\label{sec:introduction}

The origin of the most massive black holes in the Universe remains
one of the important open questions in astrophysics and cosmology.
Observations have established the existence of black holes across
an enormous range of masses, from stellar-mass objects detected
through gravitational waves~\cite{LIGOScientific:2016aoc,LIGOScientific:2018mvr},
to intermediate-mass black holes~\cite{Greene:2019vlv},
and supermassive black holes (SMBHs) residing in the centers of
galaxies~\cite{Harikane:2022rqt,CEERSTeam:2023qgy}.
Particularly intriguing is the presence of very massive black
holes at high redshifts, which challenges our understanding of
how sufficiently massive seeds can form and grow within the
available cosmic time. With black-hole masses spanning several
orders of magnitude, and reaching values as large as
$\mathcal{O}(10^{9}-10^{10}M_\odot)$ or beyond
~\cite{vanDokkum:2025bah}, identifying the physical mechanisms
responsible for their formation and subsequent evolution
remains a central task.

Primordial black holes (PBHs), which may have formed through the
gravitational collapse of enhanced density perturbations or
other processes in the early Universe
~\cite{1966AZh....43..758Z,Hawking:1971ei,Carr:1974nx},
provide an interesting possibility in this direction.
Unlike astrophysical black holes, PBHs are not restricted by
stellar evolution and can, in principle, form over a broad
range of masses~\cite{Carr:2020gox}. They have therefore been
considered both as possible constituents of dark matter and
as potential seeds of the black holes observed at later
cosmological epochs. Whether such primordial seeds can acquire
sufficiently large masses, however, depends sensitively on
their subsequent evolution. Several mechanisms have been
investigated, including direct gravitational collapse
~\cite{Zhang:2025grn}, mergers~\cite{Prole:2025snf},
and accretion from the surrounding cosmic medium
~\cite{Ziparo:2024nwh,Das:2025vts,Kalita:2025fcs}.
Among these possibilities, mass growth during the
radiation-dominated era is especially appealing, since
the early Universe contains an exceptionally dense and
hot radiation background.

Indeed, while isolated black holes lose mass through
Hawking radiation, a black hole immersed in a sufficiently
hot environment can gain energy through radiative
absorption~\cite{Meissner:2020rzo}. During the early
radiation-dominated era, the temperature of the primordial
plasma can greatly exceed the Hawking temperature of a
PBH, suggesting that absorption may dominate evaporation.
However, the existence of a favorable temperature
difference does not by itself guarantee efficient mass
growth. The actual absorption rate depends on the
microscopic properties of the surrounding radiation,
its ability to reach the black-hole horizon and the
response of the cosmological environment to the
energy extracted by the black hole.

The possibility of substantial PBH growth in the early
Universe has a long history. The pioneering analysis of
Zel'dovich and Novikov~\cite{1966AZh....43..758Z}
suggested that accretion could lead to self-similar
evolution, with the black-hole mass increasing
proportionally to cosmic time. However, Carr and
Hawking~\cite{Carr:1974nx} subsequently demonstrated
that such sustained self-similar behavior is severely
restricted for black holes embedded in an expanding
radiation universe described by a relativistic perfect
fluid. This result raises an important question, namely
to what extent can the growth of PBHs differ when
the surrounding radiation is not entirely described
by a tightly coupled fluid. In particular, after
neutrino decoupling, a significant fraction of the
radiation energy density is carried by relativistic
particles that propagate essentially collisionlessly
~\cite{Baumann:2022mni,Gonin:2026pqv}. Their capture
is governed by kinetic rather than hydrodynamic
considerations, potentially modifying the efficiency
of radiative accretion. Nevertheless, collisionless
transport alone does not remove the constraints
imposed by energy conservation and the evolution
of the surrounding spacetime.

This brings us to another fundamental issue that is
often obscured in simplified treatments of PBH
accretion. If the black-hole mass increases substantially,
the energy absorbed cannot be regarded as an
unlimited contribution from an unperturbed thermal
bath. The local radiation density must respond to
the energy transfer, while the evolving black-hole
mass modifies the surrounding gravitational geometry.
Although accretion in time-dependent cosmological
settings has been considered in various approaches
~\cite{Carr:2010wk,Babichev:2012sg,Gaur:2023hmk,Jangra:2024sif,Kalita:2025fcs}, a
consistent interpretation
of rapid mass growth requires particular care in
distinguishing the properties of the background
cosmology from those of the local black-hole
environment. In this respect, the central question
is not simply whether radiation can be absorbed,
but whether the available energy supply can sustain
significant growth without producing unphysical
runaway behavior.

In this Letter, we investigate this question by
combining radiative absorption with a description
of local radiation depletion. We first distinguish
Hawking emission, absorption from the coupled
primordial plasma and the capture of free-streaming
relativistic particles, in particular neutrinos after
their decoupling. We then use the Misner-Sharp
formalism to formulate the energy balance of a
dynamical black-hole horizon, and show that the two
absorption channels respond to depletion in opposite
ways. For free-streaming neutrinos, the ingoing flux
reaching the horizon is not depleted, because
collisionless particles crossing the horizon have
never previously interacted with the black hole.
For the coupled plasma, the energy deficit created
by accretion can only be redistributed within the
sound horizon, which therefore sets the accessible
reservoir.

With this causal depletion, we show that self-similar
growth is impossible as long as the reservoir is
smaller than $\simeq1.4$ sound horizons, in line with
the classic result of Carr and Hawking~\cite{Carr:1974nx}.
The runaway that an unperturbed bath predicts for
$f_H\simeq0.19$--$0.24$ is replaced by a finite mass
amplification, which remains below a factor of
$\simeq8$ throughout the regime where quasi-stationary
capture rates can be trusted. We also state
explicitly the assumptions this conclusion rests
on, and the regime ($f_H\gtrsim0.25$) where only a
fully dynamical treatment can be conclusive.

The paper is organized as follows. In
Section~\ref{sec:radiative} we discuss radiative
absorption and collisionless particle capture
in the early Universe. In
Section~\ref{sec:backreaction} we introduce
the Misner-Sharp energy balance, formulate the
depletion of each absorption channel and analyse
self-similar growth. The resulting mass evolution
and its cosmological implications are examined in
Section~\ref{sec:cosmological_implications},
while our conclusions are presented in
Section~\ref{sec:conclusions}.

\section{Radiative absorption in the early Universe}
\label{sec:radiative}

As Hawking showed~\cite{Hawking:1974rv}, Schwarzschild black holes emit thermal
radiation characterized by a temperature inversely proportional to their mass,
\begin{equation}
T_{\rm bh}=\frac{\hbar c^3}{8\pi G k_{\rm B}M},
\label{eq:hawking_temperature}
\end{equation}
where $M$ is the black-hole mass, $G$ is Newton's gravitational constant, $k_{\rm B}$
is Boltzmann's constant, and $c$ is the speed of light. For an isolated black hole,
Hawking emission leads to a gradual decrease of its mass, described by
\begin{equation}
\left.\frac{dM}{dt}\right|_{\rm evap}
=-\frac{\hbar c^4}{15360\pi G^2M^2}
\mathcal{G}_{\rm H}(M),
\label{eq:hawking_evaporation}
\end{equation}
where the dimensionless factor $\mathcal{G}_{\rm H}(M)$ accounts for the available
particle species and their emission greybody factors, relative to the elementary
black-body
estimate~\cite{1990PhRvD..41.3052M,MacGibbon:1991tj,Hooper:2019gtx,Arbey:2019mbc}.

However, a primordial black hole (PBH) formed during the radiation-dominated (RD) era
is not isolated, but is embedded in the hot primordial plasma. Its mass evolution is
therefore determined by the competition between Hawking emission and the absorption
of surrounding radiation. For an ideal black body, the net radiative energy transfer
is governed by the Stefan-Boltzmann law and is proportional to the difference
$T_{\rm env}^4-T_{\rm bh}^4$, where $T_{\rm env}$ denotes the environmental
temperature.
For a black hole, the corresponding rates depend additionally on the absorption
cross sections and emission greybody factors of the relevant particle species.
Nevertheless, when the surrounding radiation is much hotter than the black hole,
$T_{\rm env}\gg T_{\rm bh}$, radiative absorption can dominate Hawking evaporation.

During the RD era, the temperature of the primordial plasma evolves approximately
as~\cite{Baumann:2022mni}
\begin{equation}
T_{\rm u}=1.805\times10^8
\left(\frac{10^4\,{\rm s}}{t}\right)^{1/2}
\left(\frac{1}{g_*(T_{\rm u})}\right)^{1/4}
{\rm K},
\label{eq:cosmic_temperature}
\end{equation}
where $g_*(T_{\rm u})$ denotes the effective number of relativistic degrees of
freedom contributing to the cosmic radiation energy density. Since $T_{\rm bh}\propto
M^{-1}$, the condition $T_{\rm u}\gg T_{\rm bh}$ is readily satisfied for a broad
range of PBH masses. The corresponding absorption rate, however, depends on the
transport properties of the surrounding medium, which must be treated consistently
with its cosmological evolution.

PBHs are expected to form through the gravitational collapse of enhanced primordial
density perturbations, with initial masses related to the cosmological horizon mass
at formation~\cite{Carr:1974nx,Carr:1975qj}. More precisely, the critical-collapse
mass-scaling relation reads~\cite{Niemeyer:1997mt,Musco:2012au}
\begin{equation}
M_{\rm i}=\mathcal{K}M_{\rm H}(t_{\rm i})
(\delta-\delta_{\rm c})^{\gamma_M}
\equiv f_H M_{\rm H}(t_{\rm i}),
\label{eq:pbh_formation}
\end{equation}
where $\delta$ is the initial density contrast, $\delta_{\rm c}$ is the threshold for
PBH formation, and $\mathcal{K}$ depends on the properties of the collapsing
perturbation. The critical-collapse exponent $\gamma_M\simeq0.36$ is distinct from
the initial horizon mass fraction $f_H\equiv M_{\rm i}/M_{\rm H}(t_{\rm i})$, which
depends on the initial conditions. This distinction is essential when determining the
subsequent PBH mass evolution.

An important aspect of radiative absorption concerns the microscopic properties of
the primordial plasma. Before neutrino decoupling, photons, electrons, positrons, and
neutrinos participate in a tightly coupled relativistic medium. In this regime,
gravitational capture cannot generally be described solely through independent
particle trajectories, since interactions, pressure gradients, and the collective
response of the surrounding fluid must also be taken into account.

The situation changes after neutrino decoupling, occurring at temperatures of order
$1\,{\rm MeV}$. Neutrinos subsequently propagate approximately as collisionless,
free-streaming relativistic particles, allowing their gravitational capture to be
described through a kinetic treatment. Thus, the collisionless approximation adopted
below applies only after neutrino decoupling; the evolution of PBHs formed at earlier
times also depends on their interaction with the initially coupled plasma.
In the geometric-optics limit, the critical impact parameter for the capture of
ultrarelativistic particles by a Schwarzschild black hole is
\begin{equation}
b_{\rm c}=3\sqrt{3}\frac{GM}{c^2},
\end{equation}
yielding the effective capture cross section
\begin{equation}
\sigma_{\rm fs}=\pi b_{\rm c}^2
=\frac{27\pi G^2M^2}{c^4}
=\frac{27}{16}A_{\rm H},
\label{eq:capture_cross_section}
\end{equation}
where $A_{\rm H}=16\pi G^2M^2/c^4$ is the Schwarzschild horizon area. Thus, for a
locally isotropic distribution of ultrarelativistic free-streaming particles with
energy density $u_{\rm fs}$, the corresponding capture rate can be approximated by
\begin{equation}
\left.\frac{dM}{dt}\right|_{\rm fs}
\simeq\frac{\sigma_{\rm fs}}{c}u_{\rm fs}.
\label{eq:free_streaming_capture}
\end{equation}
This expression provides the kinetic capture rate in the local Schwarzschild
approximation, while its generalization to an evolving cosmological geometry requires
accounting for the corresponding gravitational backreaction.

After electron-positron annihilation, neutrinos and photons have different
temperatures, with $T_\nu=(4/11)^{1/3}T_{\rm u}$, where $T_{\rm u}$ now denotes
the photon temperature, in the instantaneous-decoupling approximation.
Consequently, their respective contributions to the radiation energy density
and to PBH absorption must be treated separately. In particular, the
collisionless neutrino component provides a distinct channel for gravitational
capture, whereas the photon-coupled plasma remains subject to hydrodynamic effects.
Although the photon energy density exceeds that of neutrinos after
electron-positron annihilation, the efficiency of photon absorption depends
on the collective dynamics of the coupled plasma and cannot be inferred
directly from the collisionless capture cross section.

The complete PBH mass evolution is therefore determined by the combined contributions
of Hawking evaporation, absorption from the coupled radiation plasma, and
collisionless particle capture. Crucially, the latter cannot be assumed to proceed
indefinitely in an unperturbed cosmological background: the energy absorbed by the
black hole modifies its surrounding environment and, potentially, the local spacetime
geometry. In the following section, we employ the Misner-Sharp formalism to describe
the gravitational energy balance and introduce a reduced model of local radiation
depletion to investigate its implications for PBH mass growth.

\section{Primordial black hole growth with cosmological backreaction}
\label{sec:backreaction}

The radiative absorption mechanism discussed in the previous section can lead
to an increase in the mass of primordial black holes during the
radiation-dominated era. However, estimating the resulting mass evolution
requires particular care, since the accretion of surrounding radiation
can modify both the local energy distribution and the gravitational geometry.
In particular, treating the primordial plasma as an
unperturbed thermal
bath  \cite{Haque:2026vvp,Gonin:2026pqv} becomes questionable when the mass gained
by the black hole is no
longer negligible compared to the energy available in its cosmological
environment.

To illustrate this issue, consider the approximate mass-growth equation
obtained when radiative absorption dominates Hawking evaporation and the
cosmological radiation density is assumed to follow its unperturbed
evolution,
\begin{equation}
\frac{dM}{dt}=q\frac{M^2}{t^2},
\label{eq:growth_unperturbed}
\end{equation}
where $q$ is an effective absorption coefficient determined by the
properties of the surrounding radiation. For constant $q$, integration
from an initial time $t_{\rm i}$ within the regime of applicability of
this approximation yields
\begin{equation}
M(t)=\frac{M_{\rm i}}
{1-qM_{\rm i}\left(\frac{1}{t_{\rm i}}-\frac{1}{t}\right)},
\label{eq:mass_unperturbed}
\end{equation}
where $M_{\rm i}=M(t_{\rm i})$. Depending on the initial conditions,
this expression can exhibit a strong mass amplification or even develop
a mathematical divergence. Such behavior, however, cannot be directly
interpreted as physical black-hole growth, since the calculation
neglects the response of the surrounding radiation and spacetime geometry.

Indeed, the possibility of self-similar primordial black-hole growth in
an expanding radiation universe has long been recognized as a subtle
problem. Carr and Hawking showed that sustained self-similar growth is
strongly constrained for a black hole embedded in a relativistic
perfect fluid~\cite{Carr:1974nx}. Although the collisionless particle
component discussed in Section~\ref{sec:radiative} is governed by
different transport properties, its presence does not eliminate the
requirements of local energy conservation and gravitational
self-consistency. We therefore introduce a general framework for
describing the gravitational energy of a black hole in a dynamical,
spherically symmetric spacetime, using the Misner-Sharp formalism
to characterize its interaction with the surrounding radiation.

\subsection{Misner-Sharp formalism and horizon dynamics}

A consistent description of an accreting black hole in a cosmological
environment must account for the time dependence of the geometry and
the inhomogeneous distribution of matter and radiation. In spherical
symmetry, the spacetime metric can be written in the general form
\begin{equation}
ds^2=-e^{2\Phi(t,r)}c^2dt^2
+e^{2\Lambda(t,r)}dr^2
+R^2(t,r)d\Omega^2,
\label{eq:ms_metric}
\end{equation}
where $R(t,r)$ is the areal radius and $\Phi(t,r)$ and
$\Lambda(t,r)$ characterize the dynamical geometry. This
representation allows both the gravitational field and the
surrounding energy distribution to evolve with time.

The gravitational energy contained within a sphere of areal radius
$R$ is characterized by the Misner-Sharp mass
$M_{\rm MS}(t,r)$~\cite{1964PhRv..136..571M}, defined through
\begin{equation}
1-\frac{2GM_{\rm MS}(t,r)}{c^2R}
=g^{ab}\nabla_aR\nabla_bR
=e^{-2\Lambda}(R')^2
-\frac{e^{-2\Phi}}{c^2}\dot R^2,
\label{eq:ms_mass}
\end{equation}
where a dot and a prime denote derivatives with respect to $t$
and $r$, respectively, and the indices $a,b$ refer to the
two-dimensional spacetime orthogonal to the symmetry spheres.

The black-hole apparent horizon is identified by the vanishing
of the outgoing null expansion. In spherical symmetry, its
location satisfies
\begin{equation}
g^{ab}\nabla_aR\nabla_bR\big|_{R=R_{\rm h}}=0,
\qquad
R_{\rm h}(t)=
\frac{2G M_{\rm MS}(t,r_{\rm h}(t))}{c^2}.
\label{eq:ms_horizon}
\end{equation}
Thus, the quasi-local mass evaluated at the black-hole trapping
horizon determines its instantaneous areal radius. In a
cosmological spacetime, this black-hole horizon must be
distinguished from the cosmological apparent horizon associated
with the background expansion.

The evolution of the Misner-Sharp energy is governed by the
Einstein field equations and can be expressed through the
unified first law of spherically symmetric gravitational
dynamics~\cite{Hayward:1997jp},
\begin{equation}
\nabla_a E_{\rm MS}
=A\psi_a+W\nabla_aV,
\label{eq:ms_energy_balance}
\end{equation}
where $E_{\rm MS}=M_{\rm MS}c^2$, $A=4\pi R^2$ and
$V=4\pi R^3/3$. The work density $W$ and the
energy-supply vector $\psi_a$ are defined as
\begin{equation}
W=-\frac{1}{2}T^{ab}h_{ab},
\qquad
\psi_a=T_a{}^b\nabla_bR+W\nabla_aR,
\label{eq:ms_flux}
\end{equation}
with $h_{ab}$ denoting the metric on the two-dimensional
normal spacetime and $T_{\mu\nu}$ the total
energy-momentum tensor of the surrounding matter and radiation.

Projecting Eq.~(\ref{eq:ms_energy_balance}) along the
dynamical black-hole horizon gives
\begin{equation}
\frac{d}{dt}\left(M_{\rm h}c^2\right)
=A_{\rm h}\psi_a\xi^a
+W_{\rm h}\frac{dV_{\rm h}}{dt},
\label{eq:ms_horizon_evolution}
\end{equation}
where $M_{\rm h}=M_{\rm MS}(t,r_{\rm h}(t))$,
$\xi^a=dx_{\rm h}^a/dt$ is tangent to the horizon,
and all quantities on the right-hand side are evaluated
at the horizon.

Equation~(\ref{eq:ms_horizon_evolution}) expresses the
evolution of the quasi-local horizon energy in terms of
the energy supply and work associated with the surrounding
matter. Importantly, it does not assume that the radiation
density remains homogeneous or that the background
temperature is unaffected by the energy absorbed by
the black hole. However, the Misner-Sharp identity and
the associated energy-balance relation do not, by
themselves, determine the microscopic absorption rate
or provide a closed description of the horizon evolution.

In the present context, the energy-momentum tensor must
incorporate both the coupled primordial plasma and the
collisionless relativistic component, whose capture
properties were discussed in the previous section. Their
respective contributions to the local energy flux are
generally different, and the resulting horizon evolution
must be determined consistently with the conservation
equations $\nabla_\mu T^{\mu\nu}=0$.

The Misner-Sharp framework therefore provides the
geometrical basis for investigating primordial black-hole
growth beyond the fixed-background approximation.
Nevertheless, determining the actual mass evolution
requires specifying the local radiation distribution,
its response to absorption, and the energy transport
through the surrounding environment. In the following
subsections, we introduce reduced prescriptions for
local radiation depletion and investigate their
implications for PBH growth, without solving the
complete coupled Einstein-matter system.

\subsection{Local energy depletion}
\label{sec:local_depletion}

The Misner-Sharp formulation developed above provides a covariant
description of the gravitational energy associated with a dynamical
black-hole horizon. However, determining its evolution requires
specifying how the surrounding radiation responds to the energy
transferred into the black hole. In particular, the assumption of an
unperturbed cosmological radiation density becomes inadequate when
the absorbed energy constitutes a significant fraction of the
locally available radiation. We therefore consider the effects of
local energy depletion and their implications for the subsequent
PBH mass evolution.

To characterize these effects, let us consider a spherical region
surrounding the black hole, with physical volume $V(t)$ and
radiation energy $E_{\rm r}(t)$. In a homogeneous, comoving
single-zone approximation, local energy conservation during
the radiation-dominated era gives
\begin{equation}
\frac{dE_{\rm r}}{dt}+H E_{\rm r}
=
-c^2\dot M_{\rm abs}
+\mathcal{F}_{\rm b},
\label{eq:local_energy_conservation}
\end{equation}
where $\dot M_{\rm abs}$ denotes the mass absorbed per unit time
and $\mathcal{F}_{\rm b}$ accounts for the net radiation energy
entering the region through its outer boundary. Here, the term
$H E_{\rm r}$ describes the energy loss associated with
cosmological expansion, assuming a relativistic equation of
state and $V\propto a^3$. Equation~(\ref{eq:local_energy_conservation})
is a reduced description of local energy exchange; in a
fully relativistic treatment, the corresponding flux and
work contributions are determined by
Eq.~(\ref{eq:ms_horizon_evolution}).

In the absence of absorption and boundary fluxes, the radiation
energy evolves as $E_{\rm r}^{(0)}\propto a^{-1}$. Integrating
Eq.~(\ref{eq:local_energy_conservation}) relative to this
background evolution yields
\begin{equation}
\begin{split}
E_{\rm r}(t)=E_{\rm r}^{(0)}(t)
-\frac{1}{a(t)}
\int_{t_{\rm i}}^t a(t')
\left[c^2\dot M_{\rm abs}(t')
-\mathcal{F}_{\rm b}(t')\right]dt',
\end{split}
\label{eq:radiation_depletion}
\end{equation}
where we have assumed that the local radiation energy initially
coincides with its background value. This relation shows explicitly
that radiation depletion depends not only on the mass absorbed by
the black hole but also on cosmological redshifting and the
replenishment of radiation through the outer boundary.

For the tightly coupled component of the primordial plasma,
which can be characterized by a local temperature, the radiation
energy density satisfies $u_\gamma\propto T_\gamma^4$ when the
effective relativistic degrees of freedom remain constant.
Consequently,
\begin{equation}
\frac{T_{\gamma,\rm loc}^4(t)}{T_{\rm u}^4(t)}
=
\frac{u_{\gamma,\rm loc}(t)}
{u_\gamma^{(0)}(t)}.
\label{eq:local_temperature_depletion}
\end{equation}
In contrast, after neutrino decoupling the distribution of
free-streaming neutrinos must be determined from collisionless
transport, rather than by imposing local thermal equilibrium.
The coupled plasma and the neutrino component must therefore
be evolved separately.

In the quasi-stationary Schwarzschild limit, the absorption
rates discussed in Section~\ref{sec:radiative} provide the
following approximate mass-transfer equation:
\begin{equation}
\begin{split}
\frac{dM_{\rm h}}{dt}\simeq&
\frac{16\pi G^2M_{\rm h}^2}{c^5}
\left[
\eta_\gamma u_{\gamma,\rm loc}
+\frac{27}{16}u_{\nu,\rm loc}
\right]\\
&-\frac{\hbar c^4}{15360\pi G^2M_{\rm h}^2}
\mathcal{G}_{\rm H}(M_{\rm h}),
\end{split}
\label{eq:local_mass_growth}
\end{equation}
where $M_{\rm h}$ is the Misner-Sharp mass evaluated at the
black-hole horizon, and $u_{\gamma,\rm loc}$ and $u_{\nu,\rm loc}$
are the local photon and neutrino energy densities, respectively.
The dimensionless parameter $\eta_\gamma$ describes the effective
absorption efficiency of the coupled photon plasma relative to
the horizon-area prescription, incorporating the effects of
its transport properties and collective dynamics. Unlike
collisionless neutrino capture, this contribution cannot
generally be determined from the geometric-optics cross
section alone. The factor $27/16$ corresponds to the
geometric-optics capture cross section of ultrarelativistic
collisionless particles. For the coupled plasma we adopt the
relativistic Michel-Babichev accretion rate for a perfect fluid
with $w=1/3$~\cite{Michel:1972oeq,Babichev:2004yx},
$\dot M=4\pi\lambda G^2M^2(\rho+P)/c^5$ with $\lambda=6\sqrt3$,
which corresponds to $\eta_\gamma=\lambda/3=2\sqrt3$. Equation~(\ref{eq:local_mass_growth}) represents
a local capture approximation, while the horizon dynamics
and the associated work terms must ultimately satisfy
the full Misner-Sharp energy-balance relation.

To make contact with the unperturbed growth law, we define
the local depletion factors
\begin{equation}
D_j(t)\equiv
\frac{u_{j,\rm loc}(t)}{u_j^{(0)}(t)},
\qquad j=\gamma,\nu.
\label{eq:depletion_factors}
\end{equation}
During radiation domination, the background radiation
energy density is
\begin{equation}
u_{\rm r}^{(0)}(t)=
\frac{3c^2}{32\pi Gt^2}.
\label{eq:background_radiation_density}
\end{equation}
Neglecting Hawking evaporation in the absorption-dominated
regime, Eq.~(\ref{eq:local_mass_growth}) can consequently
be expressed as
\begin{equation}
\frac{dM_{\rm h}}{dt}\simeq
q_{\rm eff}(t)
\frac{M_{\rm h}^2}{t^2},
\label{eq:effective_growth_equation}
\end{equation}
with
\begin{equation}
q_{\rm eff}(t)=
\frac{3G}{2c^3}
\left[
\eta_\gamma f_\gamma D_\gamma
+\frac{27}{16}f_\nu D_\nu
\right],
\label{eq:effective_growth_coefficient}
\end{equation}
where $f_\gamma$ and $f_\nu$ are the fractional
contributions of photons and neutrinos to the background
radiation energy density. After electron-positron
annihilation, their standard values are approximately
$f_\gamma\simeq0.595$ and $f_\nu\simeq0.405$.
These values characterize the post-annihilation radiation
background, in which neutrinos are already free-streaming.
Before neutrino decoupling, all relativistic species belong
to the coupled plasma, and
Eq.~(\ref{eq:effective_growth_coefficient}) applies with
$f_\gamma=1$, $f_\nu=0$. The coefficient $q_{\rm eff}$ is therefore
not a universal constant: it depends on the transport
properties and the local evolution of the radiation components.

A simple way to represent finite energy availability is a
fixed reservoir, $D=1-(M-M_{\rm i})/M_{\rm res}$, whose
exact solution is given in Appendix~\ref{app:analytical_solution}. $M_{\rm res}$ stands here for the mass equivalent of the initially accessible radiation reservoir. Such a prescription, however, caps the supply at the energy
initially available and ignores the transport properties
that distinguish the two channels in
Eq.~(\ref{eq:local_mass_growth}).  We therefore consider the different transport
properties of the two components and introduce corresponding depletion prescriptions.

It is convenient to write all rates in terms of the
instantaneous horizon fraction
\begin{equation}
m(t)\equiv\frac{M(t)}{M_{\rm H}(t)}=\frac{GM}{c^3t}
=\frac{r_{\rm S}}{R_{\rm H}},
\label{eq:horizon_fraction}
\end{equation}
where $M_{\rm H}(t)=c^3t/G$ and $R_{\rm H}=c/H=2ct$, so that
$m$ also measures the size of the black hole relative to
the Hubble radius. Eqs.~(\ref{eq:effective_growth_equation})
and (\ref{eq:effective_growth_coefficient}) then read
\begin{equation}
\frac{t\dot M}{M}=\frac{3}{2}\,m\left[\eta_\gamma f_\gamma D_\gamma
+\frac{27}{16}f_\nu D_\nu\right],
\label{eq:growth_m}
\end{equation}
where the subscript $\gamma$ denotes the coupled plasma
(all relativistic species before neutrino decoupling,
$f_\gamma=1$, $f_\nu=0$) and $\nu$ the free-streaming
component. In an unperturbed bath, $D_\gamma=D_\nu=1$,
Eq.~(\ref{eq:mass_unperturbed}) gives
$M_f/M_{\rm i}=(1-\hat q f_H)^{-1}$ with
$\hat q=\tfrac32(\eta_\gamma f_\gamma+\tfrac{27}{16}f_\nu)$,
i.e.\ a pole at
\begin{equation}
f_H^{\rm pole}=\hat q^{-1}=
\begin{cases}
1/(3\sqrt3)\simeq0.192, & \text{before $\nu$ decoupling},\\
0.243, & \text{after $e^\pm$ annihilation}.
\end{cases}
\label{eq:pole}
\end{equation}
The first value coincides exactly with Carr's estimate
$f_H\simeq w^{3/2}$ for $w=1/3$~\cite{Carr:1975qj}. This is
no accident: both mark the scale at which the Bondi radius
becomes comparable to the sound horizon. An unperturbed bath
therefore places PBHs formed with $f_H\gtrsim0.2$ in a
runaway regime, and it is precisely there that depletion
must be assessed.

\subsection{Depletion of the two absorption channels}

We now examine how the two radiation components respond
to the energy absorbed by the black hole. Since the coupled
plasma and free-streaming neutrinos obey different transport
equations, their depletion must be treated separately.

For collisionless neutrinos, the distribution function is
conserved along geodesics. Particles reaching the black-hole
horizon originate from the surrounding cosmological
background, while those already captured cannot return
to the external region. Consequently, the deficit in the
outgoing distribution does not affect the incoming flux
at leading order. Within the test-field and quasi-stationary
approximations, we therefore obtain
\begin{eqnarray}
\label{Ddef1}
&&
\!\!\!\!\!\!\!\!\!\!\!\!\!\!\!\!\!\!\!\!\!\!
D_\nu(t)=1+\mathcal{O}(m)+\mathcal{O}(\epsilon),
\\
&&\!\!\!\!\!\!\!\!\!\!\!\!\!\!\!\!\!\!\!\!\!\!
\epsilon\equiv\frac{G\dot M}{c^3}
=\frac{3}{2}m^2
\left[
\eta_\gamma f_\gamma D_\gamma
+\frac{27}{16}f_\nu
\right].
\end{eqnarray}
The corrections arise from the finite size of the black
hole relative to the Hubble radius, $\mathcal{O}(m)$,
and the time dependence of the geometry,
$\mathcal{O}(\epsilon)$. The neutrino capture rate can
also be compared with the growth of the neutrino energy
within the Hubble volume. Defining
$s_\nu\equiv G\dot M_\nu/(f_\nu c^3)=81m^2/32$,
we obtain $s_\nu<1$ for $m\lesssim0.63$.
Thus, the collisionless component cannot be described
by a fixed reservoir, since the incoming neutrino flux
is continuously supplied from the surrounding cosmological
region.

For the coupled plasma, absorption is instead governed
by hydrodynamic transport. The Michel--Babichev rate
assumes a stationary reservoir with asymptotic density
and pressure $(\rho_\infty,P_\infty)$ outside the sonic
radius $r_c=3GM/c^2$. In an expanding universe, however,
the energy deficit produced by accretion can only be
redistributed through pressure disturbances. During
radiation domination, the cosmological sound horizon is
\begin{equation}
R_s(t)=a(t)\int_0^t\frac{c_s\,dt'}{a(t')}
=2c_st=\frac{R_H}{\sqrt{3}}.
\end{equation}
We therefore consider a region of radius $\kappa R_s$,
with $\kappa=\mathcal{O}(1)$, as an effective radiation
reservoir. Applying the energy-balance relation (\ref{eq:radiation_depletion}),
the absorbed energy is depleted by accretion and diluted
by expansion, while the growing region allows additional
radiation to enter. The corresponding prescription is
\begin{eqnarray}
\label{Dgammaeltadef}
&&D_\gamma(t)=
\max\left[0,1-\frac{\Delta(t)}{M_s(t)}\right],
\\
&&
\Delta(t)=\frac{1}{a(t)}
\int_{t_i}^{t}a(t')\dot M_\gamma(t')\,dt',
\label{Deltadef}
\\
&&
M_s(t)=\frac{\kappa^3f_\gamma}{3\sqrt{3}}M_H(t).
\end{eqnarray}
Here, $\dot M_\gamma$ is the fluid contribution to
Eq.~(\ref{eq:growth_m}), while $(3\sqrt{3})^{-1}=(R_s/R_H)^3$
gives the fraction of the Hubble-volume energy
within the sound horizon. Since $M_s\propto t$,
the accessible reservoir grows with cosmic time,
unlike the fixed-reservoir approximation.

We emphasize that $R_s(t)$ is the cosmological sound
horizon, not the distance traveled by a disturbance
generated at PBH formation. Thus, the adopted reservoir
is a phenomenological estimate of the accessible energy,
with its spatial extent parametrized by $\kappa$.
Although this prescription incorporates depletion and
cosmological replenishment, a complete description
requires solving the coupled Einstein-fluid equations.

\subsection{Absence of self-similar growth}

We can now examine whether the above depletion prescription
allows for self-similar PBH growth, characterized by a constant
horizon mass fraction $m$, or equivalently $t\dot M/M=1$.
During radiation domination, $a\propto t^{1/2}$, and for
$\dot M_\gamma={\rm const}$, Eq.~(\ref{Deltadef}) gives
$\Delta\simeq(2/3)t\dot M_\gamma$ at $t\gg t_i$.
The corresponding depletion factor becomes
\begin{equation}
\begin{split}
&D_\gamma=1-\frac{3\sqrt3\,\eta_\gamma}{\kappa^3}
m^2D_\gamma\\
&\Longrightarrow\quad
D_\gamma^{\rm ss}=\frac{1}{1+18m^2/\kappa^3},
\end{split}
\label{eq:D_selfsimilar}
\end{equation}
independently of $f_\gamma$. For a purely fluid radiation
bath, substituting Eq.~(\ref{eq:D_selfsimilar}) into
Eq.~(\ref{eq:growth_m}) and imposing $t\dot M/M=1$
yields
\begin{equation}
\frac{18}{\kappa^3}m^2-3\sqrt3\,m+1=0.
\label{eq:selfsimilar_condition}
\end{equation}
Interestingly, this equation admits real solutions only
for $\kappa^3\geq8/3$. Therefore, when
\begin{equation}
\kappa<\kappa_c=(8/3)^{1/3}\simeq1.39,
\label{eq:kappa_c}
\end{equation}
sustained self-similar growth is not possible within the
present model, and the horizon mass fraction eventually
decreases. This behavior is consistent with the classical
result of Carr and Hawking~\cite{Carr:1974nx}, although
our conclusion follows from the adopted depletion
prescription rather than from a fully relativistic
treatment.

For $\kappa>\kappa_c$, Eq.~(\ref{eq:selfsimilar_condition})
admits two roots, $m_\pm$, with the lower one being
unstable and defining a threshold for self-similar
evolution. For instance, taking $\kappa=\sqrt3$,
corresponding to a reservoir extending to the Hubble
radius, gives $m_-\simeq0.23$, while the second root,
$m_+\simeq1.27$, lies outside the physically relevant
subhorizon regime. After electron-positron annihilation,
the free-streaming neutrino contribution modifies the
growth condition by an additional term
$(81/32)f_\nu m$. Restricting to $m\leq0.4$, we obtain
$\kappa_c\simeq1.38$, close to the purely fluid result.
A self-similar branch may reappear at $m\gtrsim0.8$,
where, however, $s_\nu>1$ and the test-field
approximation is no longer applicable. Thus, within
the regime of validity of our treatment, a reservoir
of approximately the sound-horizon size does not
support sustained self-similar PBH growth.

\section{Cosmological implications}
\label{sec:cosmological_implications}

\begin{figure*}[t]
\centering
\includegraphics[width=\textwidth]{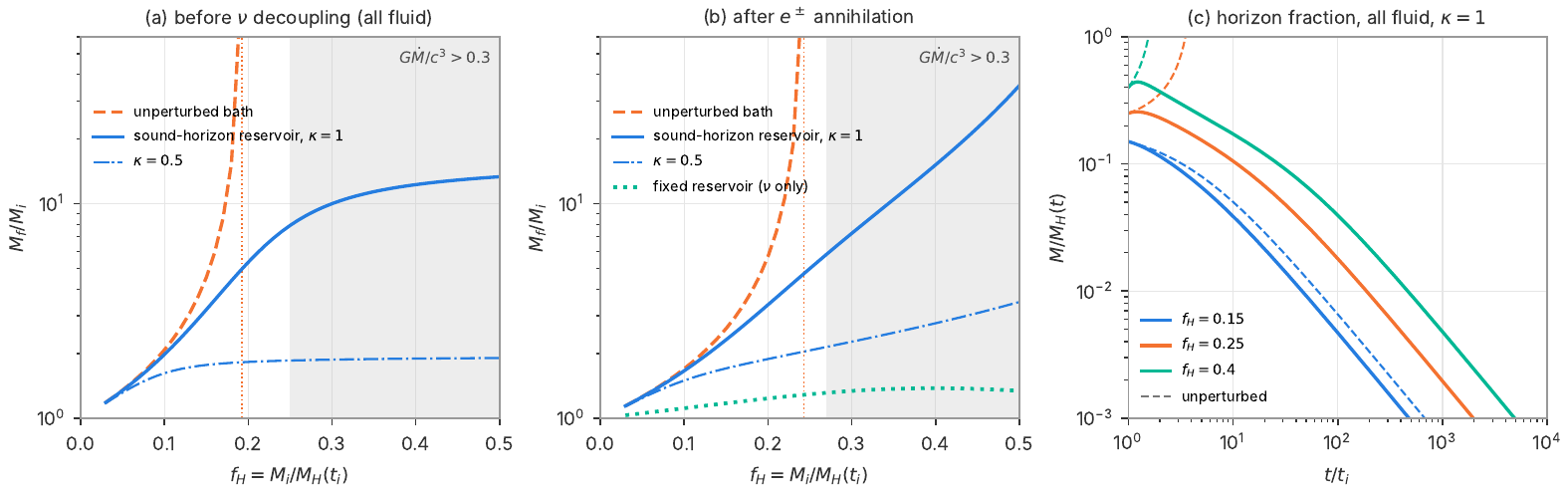}
\caption{{\it{Asymptotic PBH mass amplification
$M_f/M_{\rm i}$ as a function of the initial horizon
mass fraction $f_H$, before neutrino decoupling (a)
and after electron-positron annihilation (b).
Dashed curves correspond to the unperturbed radiation
background, with the vertical dotted lines indicating
the poles of Eq.~(\ref{eq:pole}). Solid and dash-dotted
curves show the sound-horizon depletion model for
$\kappa=1$ and $\kappa=0.5$, respectively.
In panel (b), the dotted curve represents the
fixed-reservoir, neutrino-only approximation of
Appendix~\ref{app:analytical_solution}, with
$M_{\rm res}=f_\nu[M_{\rm H}(t_{\rm i})-M_{\rm i}]$.
Shaded regions indicate $\epsilon_{\max}>0.3$,
where the quasi-stationary approximation becomes
unreliable. Panel (c) shows the evolution of the
horizon mass fraction $m=M/M_{\rm H}(t)$ for
different initial values of $f_H$ in the purely
fluid case, comparing sound-horizon depletion
with $\kappa=1$ (solid) and the corresponding
unperturbed solutions (dashed).}}}
\label{fig:growth}
\end{figure*}

\begin{figure}[t]
\centering
\includegraphics[width=\columnwidth]{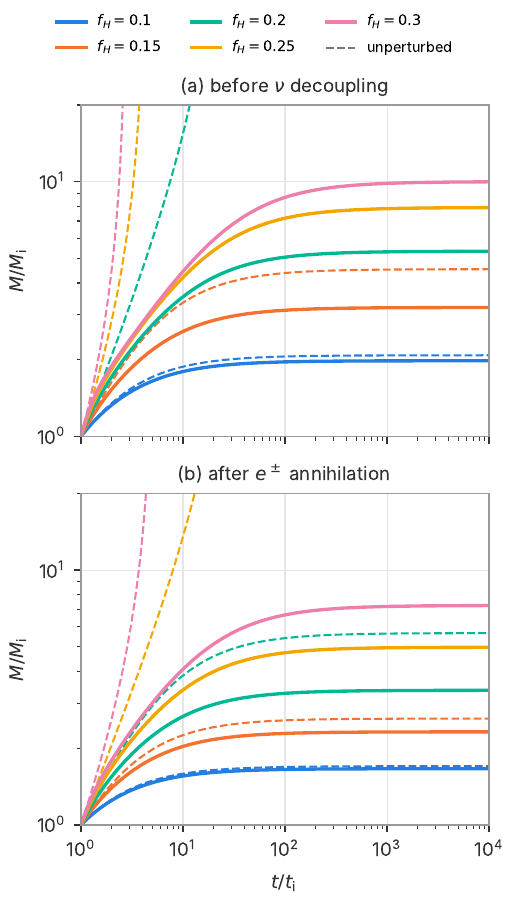}
\caption{{\it{Evolution of the PBH mass ratio
$M(t)/M_{\rm i}$ for different initial horizon
mass fractions $f_H$, before neutrino decoupling
(a) and after electron-positron annihilation (b).
Solid curves correspond to sound-horizon depletion
with $\kappa=1$, while dashed curves show the
unperturbed-background approximation, which
develops a finite-time divergence above the
critical values of Eq.~(\ref{eq:pole}).
Within the plotted range, the inclusion of local
depletion leads to a finite asymptotic mass,
with $90\%$ of the total mass increase attained
within approximately $15$--$160\,t_{\rm i}$.}}}
\label{fig:mass_evolution}
\end{figure}

We now investigate the cosmological implications of the
preceding analysis by numerically integrating
Eqs.~(\ref{eq:growth_m}), (\ref{Ddef1}), (\ref{Dgammaeltadef}) and
(\ref{Deltadef}), taking $D_\nu=1$ and imposing
$M(t_{\rm i})=f_HM_{\rm H}(t_{\rm i})$.
We adopt $\eta_\gamma=2\sqrt3$ and neglect Hawking
evaporation, focusing on the regime where radiative
absorption dominates, $T_{\rm bh}\ll T_{\rm u}$.
We consider separately two epochs with approximately
constant radiation composition, namely before neutrino
decoupling, when $f_\gamma=1$ and $f_\nu=0$, and after
electron-positron annihilation, when $f_\gamma=0.595$
and $f_\nu=0.405$. The corresponding results for
$\kappa=1$ are summarized in Table~\ref{tab:results},
while the mass amplification and its time evolution
are presented in Figs.~\ref{fig:growth} and
\ref{fig:mass_evolution}, respectively.

\begin{table}[t]
\centering
\begin{tabular}{c|cc|cc|cc}
\hline
 & \multicolumn{2}{c|}{unperturbed}
 & \multicolumn{2}{c|}{$\kappa=1$: $M_f/M_{\rm i}$}
 & \multicolumn{2}{c}{$\kappa=1$: $\epsilon_{\max}$}\\
$f_H$ & pre & post & pre & post & pre & post\\
\hline
0.10 & 2.08 & 1.70 & 1.98 & 1.66 & 0.05 & 0.04\\
0.15 & 4.53 & 2.61 & 3.21 & 2.32 & 0.12 & 0.09\\
0.20 & $\infty$ & 5.66 & 5.33 & 3.38 & 0.21 & 0.17\\
0.25 & $\infty$ & $\infty$ & 7.92 & 4.98 & 0.33 & 0.26\\
0.30 & $\infty$ & $\infty$ & 10.0 & 7.27 & 0.47 & 0.37\\
0.40 & $\infty$ & $\infty$ & 12.3 & 15.2 & 0.83 & 0.66\\
\hline
\end{tabular}
\caption{{\it{Asymptotic PBH mass amplification before
neutrino decoupling (``pre'', all fluid) and after
electron-positron annihilation (``post''), for an
unperturbed radiation background and for sound-horizon
depletion with $\kappa=1$. The symbol $\infty$ denotes
the finite-time divergence of the unperturbed solution.
The last two columns give the maximum value of
$\epsilon=G\dot M/c^3$ along each trajectory. Results
with $\epsilon_{\max}\gtrsim0.3$ lie outside the
regime of validity of the quasi-stationary
approximation and should be regarded as indicative.}}}
\label{tab:results}
\end{table}

As shown in Fig.~\ref{fig:mass_evolution}, the mass
evolution differs substantially from that obtained
in an unperturbed radiation background. In the latter
case, the solution develops a finite-time divergence
for initial fractions exceeding the critical values
$f_H\simeq0.192$ before neutrino decoupling and
$f_H\simeq0.243$ after electron-positron annihilation.
When the depletion of the coupled plasma is included,
the corresponding trajectories instead approach
finite asymptotic masses over the parameter range
considered. For the initial fractions displayed in
Fig.~\ref{fig:mass_evolution}, most of the growth
occurs within the first few hundred initial times.
The suppression is already significant below the
unperturbed pole. For instance, after electron-positron
annihilation, taking $f_H=0.2$ reduces the final mass
amplification from approximately $5.7$ to $3.4$.

The dependence on the initial horizon fraction is
shown more explicitly in Figs.~\ref{fig:growth}(a)
and (b). Within the range where the quasi-stationary
approximation remains applicable, the strong growth
associated with the pole of the unperturbed solution
is replaced by a moderate amplification, reaching
values of a few and approaching a factor of eight
near the boundary of validity. Furthermore, the
evolution of the horizon mass fraction, presented
in Fig.~\ref{fig:growth}(c), shows that $m(t)$
remains close to its initial value, with
$m_{\max}\lesssim1.1f_H$ for the illustrated
all-fluid trajectories, and subsequently decreases
approximately as $t^{-1}$. For smaller initial
fractions, $f_H\lesssim0.15$, the depletion correction
remains below approximately $30\%$, and the
unperturbed approximation provides a reasonable
description of the mass evolution.

The comparison between the two cosmological epochs
also reveals the different roles of hydrodynamic
accretion and collisionless capture. For
$f_H\lesssim0.3$, the mass amplification is smaller
after electron-positron annihilation, since part
of the coupled radiation fluid is replaced by
free-streaming neutrinos, whose effective capture
coefficient per unit energy density, $27/16$, is
smaller than the fluid coefficient
$\eta_\gamma=2\sqrt3$. At larger initial fractions,
the ordering can be reversed, as the neutrino
component remains undepleted at leading order
while the fluid contribution becomes progressively
suppressed. However, this behavior occurs mainly
in the shaded region of Fig.~\ref{fig:growth}(b),
where $\epsilon_{\max}>0.3$ and the quantitative
predictions of the quasi-stationary approximation
are no longer reliable.

To examine the dependence on the available radiation
energy, we also consider a smaller reservoir,
$\kappa=0.5$, shown by the dash-dotted curves in
Figs.~\ref{fig:growth}(a) and (b). In this case,
depletion is considerably stronger, with
$M_f/M_{\rm i}\lesssim1.9$ before neutrino decoupling
and $\lesssim2.8$ after annihilation, for
$f_H\leq0.4$. Thus, within the adopted prescription,
reducing the accessible reservoir further suppresses
the mass growth, while the results for $\kappa=1$
provide an upper estimate for smaller reservoirs.
For comparison, Fig.~\ref{fig:growth}(b) also includes
the fixed-reservoir, neutrino-only approximation of
Appendix~\ref{app:analytical_solution}. Its maximum
amplification, $M_f/M_{\rm i}\simeq1.38$, is smaller
than that of the sound-horizon model over the plotted
range, since it neglects accretion from the coupled
plasma and restricts neutrino capture to a fixed
initial energy budget.

The applicability of these results depends on both
the cosmological epoch and the assumptions underlying
the accretion rates. Since most of the growth occurs
within a few hundred $t_{\rm i}$, the constant-composition
approximation is appropriate for PBHs formed sufficiently
before neutrino decoupling,
$t_{\rm i}\lesssim3\times10^{-3}\,{\rm s}$, or after
electron-positron annihilation. PBHs formed during the
intermediate epoch require following the evolution
of the radiation composition and the corresponding
capture rates.

An additional uncertainty concerns the size of the
accessible radiation reservoir. As shown in
Section III.D, sustained self-similar growth is
excluded within the present closure for
$\kappa<\kappa_c\simeq1.39$, whereas extending
the reservoir to the Hubble radius,
$\kappa=\sqrt3$, allows a self-similar threshold
near $m_-\simeq0.23$. The choice $\kappa\simeq1$
is motivated by the finite propagation speed of
pressure perturbations, although the surrounding
underdense region associated with PBH formation
may further reduce the effective reservoir.
More importantly, the quasi-stationary treatment
requires both $m\ll1$ and $\epsilon\ll1$.
For initial fractions approaching and exceeding
$f_H\simeq0.25$, these conditions become
increasingly difficult to satisfy. The corresponding
results in Table~\ref{tab:results} and the figures
should therefore be regarded as illustrative,
rather than as reliable predictions of fully
relativistic PBH evolution. Furthermore, initial
mass fractions inferred from numerical-relativity
simulations may already incorporate early
post-formation accretion, which should not be
counted again.

These considerations are particularly relevant
for the possible formation of supermassive black
holes. Producing an object of mass
$\mathcal{O}(10^9M_\odot)$ from a substantially
lighter primordial seed would require mass
amplification by many orders of magnitude,
whereas the present analysis predicts only
moderate growth within its regime of validity.
Moreover, the free-streaming neutrino channel
becomes relevant only after neutrino decoupling.
PBHs forming at $t_{\rm i}\gtrsim1\,{\rm s}$
already have initial masses of order
$M_{\rm i}\gtrsim2\times10^5f_HM_\odot$,
while PBHs formed sufficiently earlier
generally have a much smaller horizon mass
fraction by the time neutrinos decouple.
Consequently, the present mechanism does
not provide evidence that radiative absorption
during the radiation-dominated era can
transform light primordial seeds into
supermassive black holes.

Finally, the cosmological abundance of PBHs
introduces an additional consistency requirement.
For a population described by a mass function
$dn_{\rm PBH}/dM_{\rm i}$, the energy transferred
from radiation into black holes must be compatible
with the evolution of the total cosmological
energy density. Even a moderate amplification
increases the PBH energy density at fixed
comoving number density, which must be
taken into account when comparing the
formation abundance with observational
constraints. A complete determination of
the resulting PBH mass spectrum therefore
requires evolving both the individual mass
trajectories and their cosmological abundance.

\section{Conclusions}
\label{sec:conclusions}

In this work, we have investigated the growth of primordial
black holes through radiative absorption during the
radiation-dominated era, distinguishing the hydrodynamic
accretion of the coupled primordial plasma from the
collisionless capture of free-streaming neutrinos. Using
the Misner-Sharp framework to describe the gravitational
energy balance, we have introduced a phenomenological
prescription for local radiation depletion and examined
its implications for PBH mass evolution.

We have shown that treating the primordial radiation as
an unperturbed background can lead to unphysical runaway
growth for initial horizon mass fractions
$f_H\gtrsim0.19$ before neutrino decoupling and
$f_H\gtrsim0.24$ after electron-positron annihilation.
However, the two absorption channels respond differently
to the energy extracted by the black hole. While the
incoming flux of collisionless neutrinos remains
undepleted at leading order, the energy deficit produced
by fluid accretion can only be redistributed within a
causally accessible region determined by the sound horizon.
Within the adopted depletion model, sustained self-similar
growth is not possible for reservoirs smaller than
approximately $\kappa_c\simeq1.39$ sound horizons, in
qualitative agreement with the classical conclusions
of Carr and Hawking~\cite{Carr:1974nx}.

Our numerical analysis further shows that, for a
sound-horizon-sized reservoir, the runaway behavior of
the unperturbed approximation is replaced by finite
mass amplification over the physically relevant parameter
range. The resulting growth is typically limited to
a factor of a few, approaching $\simeq8$ only near the
boundary of validity of the quasi-stationary approximation.
These results highlight the importance of accounting
for both local depletion and the transport properties
of the surrounding radiation, while demonstrating that
a fixed-reservoir approximation can underestimate the
growth by neglecting cosmological replenishment.

Consequently, our analysis does not support the formation
of supermassive black holes from substantially lighter
primordial seeds through radiative absorption alone.
Although larger initial horizon fractions may allow
different behavior, the corresponding evolution lies
beyond the regime where quasi-stationary capture rates
can be reliably applied. A definitive assessment of
such configurations requires a fully dynamical treatment
of the coupled black-hole geometry and surrounding
radiation, including the effects of gravitational
backreaction and energy transport.

{\bf{Acknowledgments}}
TP would like to thank Dimitrios Kranas for exchanges regarding the conditions on the PBH mass for a dominant radiative absorption regime.
DK is also grateful to Theocharis Apostolatos for valuable discussions regarding the manuscript.
The authors acknowledge  the  support from the COST
Actions CA21136 -  Addressing observational tensions in cosmology with
systematics and fundamental physics (CosmoVerse)  - CA23130, Bridging high and
low energies in search of quantum gravity (BridgeQG) and CA21106 -- COSMIC
WISPers in the Dark Universe: Theory, astrophysics and experiments
(CosmicWISPers). TP acknowledges as well financial support by the funding program
“MEDICUS” of the University of Patras.

\appendix

\section{Fixed-reservoir limit}
\label{app:analytical_solution}

In this Appendix, we provide the analytical solution for the
fixed-reservoir prescription mentioned in
Section~\ref{sec:local_depletion}, which is shown for comparison
in Fig.~\ref{fig:growth}(b). Within this approximation, the
mass-growth equation takes the form
\begin{equation}
\frac{dM}{dt}
=
q\frac{M^2}{t^2}
\left(1-\frac{M-M_{\rm i}}{M_{\rm res}}\right),
\label{eq:app_growth}
\end{equation}
where $q>0$ is the effective absorption coefficient,
$M_{\rm i}$ is the initial black-hole mass, and
$M_{\rm res}>0$ represents the mass equivalent of the
initially accessible radiation reservoir.

Introducing $K=M_{\rm i}+M_{\rm res}$, Eq.~(\ref{eq:app_growth})
can be rewritten as
\begin{equation}
\frac{dM}{dt}
=
\frac{q}{M_{\rm res}}
\frac{M^2(K-M)}{t^2}.
\label{eq:app_growth_recast}
\end{equation}
The equation is separable, yielding
\begin{equation}
\frac{M_{\rm res}\,dM}{M^2(K-M)}
=
q\frac{dt}{t^2}.
\label{eq:app_separable}
\end{equation}
Using the partial-fraction decomposition
\begin{equation}
\frac{1}{M^2(K-M)}
=
\frac{1}{K^2}
\left(\frac{1}{M}+\frac{1}{K-M}\right)
+\frac{1}{KM^2},
\label{eq:app_partial_fractions}
\end{equation}
and integrating with the initial condition
$M(t_{\rm i})=M_{\rm i}$, we obtain
\begin{equation}
\begin{split}
&\ln\left[
\frac{M(t)}{K-M(t)}
\frac{K-M_{\rm i}}{M_{\rm i}}
\right]
-K\left(\frac{1}{M(t)}-\frac{1}{M_{\rm i}}\right)
\\
&\hspace{1cm}=
\frac{qK^2}{M_{\rm res}}
\left(\frac{1}{t_{\rm i}}-\frac{1}{t}\right).
\end{split}
\label{eq:app_implicit_solution}
\end{equation}

The solution can be expressed explicitly through the
Lambert function $W_0$, defined by
$W_0(x)e^{W_0(x)}=x$ for $x\geq0$.
Introducing the variable
\begin{equation}
z(t)=\frac{K}{M(t)}-1,
\qquad
z_{\rm i}=\frac{M_{\rm res}}{M_{\rm i}},
\label{eq:app_variable}
\end{equation}
Eq.~(\ref{eq:app_implicit_solution}) becomes
\begin{equation}
z(t)e^{z(t)}
=
z_{\rm i}
\exp\left[
z_{\rm i}
-\frac{qK^2}{M_{\rm res}}
\left(\frac{1}{t_{\rm i}}-\frac{1}{t}\right)
\right].
\label{eq:app_lambert}
\end{equation}
Consequently, the exact mass evolution within this
approximation is
\begin{equation}
M(t)=
\frac{K}{
1+W_0\left[
z_{\rm i}
\exp\left(
z_{\rm i}
-\frac{qK^2}{M_{\rm res}}
\left(\frac{1}{t_{\rm i}}-\frac{1}{t}\right)
\right)
\right]}.
\label{eq:app_exact_solution}
\end{equation}

For $t\geq t_{\rm i}$, the argument of $W_0$ is
positive, and the solution remains regular.
Moreover, Eq.~(\ref{eq:app_growth_recast}) implies
$dM/dt>0$ for $M_{\rm i}<M<K$, while the growth
rate vanishes as $M\rightarrow K$. Therefore,
\begin{equation}
M_{\rm i}\leq M(t)<M_{\rm i}+M_{\rm res}.
\label{eq:app_mass_bound}
\end{equation}
The asymptotic mass is finite and is given by
\begin{equation}
M_f=
\frac{K}{
1+W_0\left[
z_{\rm i}
\exp\left(
z_{\rm i}-\frac{qK^2}{M_{\rm res}t_{\rm i}}
\right)
\right]}.
\label{eq:app_final_mass}
\end{equation}

Unlike the corresponding fixed-background equation,
the finite-reservoir model cannot develop a
finite-time mass divergence. This regularization
follows directly from the depletion factor in
Eq.~(\ref{eq:app_growth}), which reduces the
absorption rate as the available radiation is
consumed. Nevertheless, the result applies to
the prescribed finite-reservoir approximation;
it does not constitute an exact solution of the
coupled Einstein-matter system. In particular,
the effects of radiation replenishment and
spacetime evolution must be determined separately.

\bibliographystyle{JHEP}
\bibliography{bibliography}

\end{document}